\documentclass[sloppy,11pt]{article}

\usepackage{amsfonts,amssymb,amsthm,amsmath}
\usepackage{hyperref}
\usepackage{mathtools}
\usepackage{comment}

\usepackage{newpxtext} 
\usepackage{textcomp} 
\usepackage[varg,bigdelims]{newpxmath}
\usepackage[margin=1in]{geometry}

\usepackage[utf8]{inputenc} 
\usepackage{amsmath,amssymb,amsthm}
\usepackage{algorithm}

\usepackage{float}
\usepackage{verbatim}

\usepackage{todonotes}

\DeclareMathOperator {\col}{Col}

\usepackage[final]{listings}
\lstdefinelanguage{pseudo}{%
  morekeywords={Input,Output,if,then,else,elsif,endif,while,loop,do,done,repeat,until,for,all,to,downto,do,end,break,and,or,fail,return,output,input,true,false,guess,randomly,accept,reject,goto,halt,div,mod,case},
  sensitive=false,
  morecomment=[l]{\\},
  morecomment=[s]{/*}{*/},
  mathescape=true,
  escapeinside={\%}{\%},
}

\title{Matrix Polynomial Factorization via Higman Linearization}

\author{V. Arvind\thanks{Institute of Mathematical Sciences, Chennai,
    India, \texttt{email: arvind@imsc.res.in}} \and Pushkar S
  Joglekar\thanks{Vishwakarma Institute of Technology, Pune, India,
    \texttt{email: joglekar.pushkar@gmail.com}}
    \thanks{Author would like to thank SERB for the funding through the MATRICS project, File no. MTR/2018/001214}
    }

\date{}

\newtheorem{theorem}{Theorem}[section]

\newtheorem{definition}[theorem]{Definition}
\newtheorem{lemma}[theorem]{Lemma}

\newtheorem{claim}[theorem]{Claim}
\newtheorem{remark}[theorem]{Remark}

\newenvironment{proofof}[1]{\noindent{\it Proof of #1. }} {{\qed}}

\newtheorem*{theorem*}{Theorem}
\newtheorem*{corollary*}{Corollary}

\usepackage{enumitem}

\def\qed{\hspace*{\fill} $\Box$\par\medskip}

\usepackage{graphicx}

\usepackage{amsmath,amssymb}

\newcommand{\F}{\mathbb{F}}
\renewcommand{\angle}[1]{\langle #1 \rangle}
\newcommand{\FX}{\F\angle{X}}

\newcommand{\bit}{\mathfrak{b}}

\newcommand{\poly}{\mathrm{poly}}

\newcommand{\Q}{\mathbb{Q}}

\newcommand{\fR}{\FX}

\DeclareFontEncoding{LS1}{}{}
\DeclareFontSubstitution{LS1}{stix}{m}{n}
\DeclareSymbolFont{symbols2stix}{LS1}{stixfrak} {m} {n}
\DeclareMathSymbol{\lparenless}{\mathopen} {symbols2stix}{"32}
\DeclareMathSymbol{\rparengtr}{\mathclose}{symbols2stix}{"33}

\usepackage{framed}

\begin{document}

\maketitle

\begin{abstract}  
 In continuation to our recent work \cite{AJ22} on noncommutative
 polynomial factorization, we consider the factorization problem for
 matrices of polynomials and show the following results.
\begin{itemize}  
\item Given as input a full rank $d\times d$ matrix $M$ whose entries
  $M_{ij}$ are polynomials in the free noncommutative ring
  $\F_q\angle{x_1,x_2,\ldots,x_n}$, where each $M_{ij}$ is given by a
  noncommutative arithmetic formula of size at most $s$, we give a
  randomized algorithm that runs in time polynomial in $d,s, n$ and
  $\log_2q$ that computes a factorization of $M$ as a matrix product
  $M=M_1M_2\cdots M_r$, where each $d\times d$ matrix factor $M_i$ is
  irreducible (in a well-defined sense) and the entries of each $M_i$
  are polynomials in $\F_q\angle{x_1,x_2,\ldots,x_n}$ that are output
  as algebraic branching programs. We also obtain a deterministic
  algorithm for the problem that runs in $\poly(d,n,s,q)$.
\item A special case is the efficient factorization of matrices whose
  entries are univariate polynomials in $\F[x]$. When $\F$ is a finite
  field the above result applies. When $\F$ is the field of rationals
  we obtain a deterministic polynomial-time algorithm for the problem.
  \end{itemize}


\noindent\textbf{Keywords: Noncommutative Polynomials, Arithmetic
  Circuits, Factorization, Identity testing.}
\end{abstract}

\section{Introduction}\label{intro}

Let $\fR$ denote the free noncommutative polynomial ring in variables
$X=\{x_1,x_2,\ldots,x_n\}$ over a field $\F$. The elements of $\fR$
are \emph{noncommutative polynomials}: finite $\F$-linear combinations
of monomials (words) over the variables $X$.

\begin{definition}[Matrix Polynomials]
For a positive integer $d$, a $d\times d$ matrix $M\in\fR^{d\times d}$ over
the noncommutative polynomial ring $\fR$ is a \emph{matrix
  polynomial}. Equivalently, we can consider $M$ as an element of the
ring $\F^{d\times d}\angle{X}$ of noncommutative polynomials whose
coefficients are from the scalar matrix ring $\F^{d\times d}$.
\end{definition}

In this paper we study the problem of factorization of matrix
polynomials in $\fR^{d\times d}$ with the aim of designing
efficient algorithms. To the best of our knowledge, this is the first
algorithmic study of the problem from the viewpoint of obtaining
polynomial-time algorithms for it.

The mathematics that underlies such factorizations in noncommutative
rings is a general theory of the so-called free ideal rings due to
Cohn \cite{Cohnfir}. The matrix ring $\fR^{d\times d}$ for $d>1$,
unlike the polynomial ring $\fR$ itself, is not a \emph{domain}
because it contains zero divisors (and even nilpotent
elements). However, Cohn's factorization theory applies to \emph{full
  matrices}: a matrix $M\in\fR^{d\times d}$ is full if it cannot be
expressed as a product
\[
M= A B
\]
of matrices $A\in\fR^{d\times r}$ and $B\in\fR^{r\times d}$, where
$r<d$. In other words, a full matrix in $\fR^{d\times d}$ has
\emph{noncommutative rank} $d$. We note here that the problem of
computing noncommutative rank has received a lot of attention in
recent years \cite{GGOW20,IQS18}.

A matrix $U\in\fR^{d\times d}$ is a \emph{unit} if it is invertible in
$\fR^{d\times d}$. That is, there is a matrix $U'\in\fR^{d\times d}$
such that $UU'=U'U=I_d$. Analogous to the usual setting of commutative
unique factorization domains, we are interested in the factorization
of \emph{non-units} that are full matrices in $\fR^{d\times d}$. A
full non-unit matrix $M\in\fR^{d\times d}$ is an \emph{atom} if $M=AB$
implies either $A$ or $B$ is a unit.\footnote{Following Cohn
  \cite{Cohnfir} we refer to such matrices as atoms rather than
  irreducibles.}

Elements $A \in \fR^{d \times d}$ and $B \in \fR^{d' \times d'}$ are
called \emph{stable associates} or simply associates if there are positive integers $t$ and
$t'$ such that $d+t=d'+t'$ and units $P, Q\in\fR^{(d+t)\times (d+t)}$
such that $A\oplus I_t=P(B \oplus I_{t'})Q$. Notice that if $A$ and $B$ are full non-unit matrices that are stable
associates then $A$ is atom if and only if $B$ is atom \cite{Cohnfir}.  

Let $M\in\fR^{d\times d}$ be a full non-unit matrix polynomial. By a
\emph{complete factorization}\footnote{$\fR^{d\times d}$ is not a UFD in the usual sense, nevertheless factorization in $\fR^{d\times d}$ is unique under stable associativity as shown in \cite{Cohnfir}. However, in the current work we are not interested in this aspect, our goal is simply to find \emph{any} factorization of given matrix polynomial into atoms} of $M$ we mean expressing $M$ as a
product of matrix polynomials
\[
M= M_1 M_2\cdots M_r,
\]
where each $M_i\in\fR^{d\times d}$ is an atom.

As the first result in this paper, we consider the factorization of
matrix polynomials $M\in\fR^{d\times d}$ over a finite field $\F=\F_q$
where the entries of $M$ are given as input by noncommutative
arithmetic formulas of size $s$ and we obtain a $\poly(d,s,\log_2q)$
time randomized algorithm (we also obtain a deterministic
$\poly(d,s,q)$ time). Indeed, the only place where we require
randomness is for univariate polynomial factorization over large
characteristic fields $\F_q$.

Unfortunately, we do not have a similar result for matrix polynomials
when the underlying field $\F=\Q$ because the above method is based on
Ronyai's algorithm for computing a nontrivial common invariant
subspace for a collection of matrices which is not known to have an
efficient algorithm over rationals \cite{Ron87,FR85}. However, the
approach works for factorization of matrix polynomials over the
univariate ring $\Q[x]$.

Before we proceed we recall some basic definitions. Let $\F$ be any
field and $X=\{x_1,x_2,\ldots,x_n\}$ be a set of $n$ free noncommuting
variables. Let $X^*$ denote the set of all free words (which are
monomials) over the alphabet $X$ with concatenation of words as the
monoid operation and the empty word $\epsilon$ as identity element.
The \emph{free noncommutative ring} $\FX$ consists of all finite
$\F$-linear combinations of monomials in $X^*$, where the ring
addition $+$ is coefficient-wise addition and the ring multiplication
$*$ is the usual convolution product.

For $f\in\FX$ let $f(m)\in\F$ denote the coefficient of monomial $m$
in $f$. We can write $f=\sum_m f(m) m$. The \emph{degree} of a
monomial $m\in X^*$ is its length, and the degree $\deg f$ of a
polynomial $f\in\FX$ is the degree of a largest degree monomial in $f$
with nonzero coefficient. 




For further required background on Cohn's factorization theory we
refer the reader to our recent arxiv paper \cite{AJ22} on which the
present work is based. Cohn's texts \cite{Cohnfir,Cohnintro} contain a
comprehensive treatment.

\subsection{Overview of the results}

We show the following results.

\begin{enumerate}
\item Given a full non-unit matrix polynomial
  $M\in\F_q\angle{X}^{d\times d}$, where each of its entries $M_{ij}$
  is given as input by a noncommutative arithmetic formula of size
  $s$, a factorization $M=M_1M_2\cdots M_r$ can be computed in
  randomized time $\poly(s,\log_2q,|X|)$, where each $M_i \in \F_q \angle{X}^{d \times d}$ is an atom
  whose entries are output as algebraic branching programs (of size
  $\poly(s,\log_2q,|X|)$). We also obtain a deterministic $\poly(s,q,|X|)$
  time algorithm for the problem. 
\item For a univariate matrix polynomial $M\in\Q[x]^{d\times d}$ that
  is a full non-unit matrix we give a deterministic time $\poly(d,s)$
  time algorithm to compute factorization $M=M_1M_2\cdots M_r$ where each $M_i \in \Q[x]^{d \times d}$ is an atom and $s$ bounds the bit complexity of the rational
  coefficients of the matrix entries $M_{ij}\in\Q[x], 1\le i,j\le d$. \end{enumerate}

To the best of our knowledge, these are the first algorithms with
polynomially bounded running time for the above problems.

\begin{remark}
  We note that for univariate matrix polynomials over finite fields
  the first result yields a randomized $\poly(s,\log_2q,|X|)$ time
  algorithm and a deterministic $\poly(s,q,|X|)$ time
  algorithm. However, over rationals we do not have an analogous
  efficient algorithm for factorizing matrix polynomials in
  $\Q\angle{X}^{d\times d}$ ever for $d=1$. As explained in
  \cite{AJ22} the approach to factorization we use breaks down over
  $\Q$ because the problem of computing nontrivial common invariant
  subspaces of a collection of rational matrices is at least as hard
  as factoring square-free integers \cite{FR85}.
\end{remark}  

The algorithm for matrix polynomial factorization uses the same
strategy as we did for noncommutative polynomial factorization
\cite{AJ22}. We briefly outline it.

\begin{itemize}
\item \textbf{Higman linearization}~~ Given a non-unit matrix polynomial
  $M\in\FX^{d\times d}$, where each entry is input by a noncommutative
  formula, we transform it into a linear matrix $L$ such that $M\oplus
  I = PLQ$, where $P,Q,L\in\FX^{r\times r}$ for $r=O(d^2s)$, $P$ is
  upper triangular with all $1$'s diagonal, $Q$ is lower triangular
  with all $1$'s diagonal, and $L$ is a full non-unit linear
  matrix. 

\item \textbf{Linear Matrix Factorization}~~ Next, we factorize the
  linear matrix $L$ as a product of atomic linear factors using our
  algorithm described in \cite{AJ22}.  This algorithm is based on
  ideas from Cohn's factorization theory \cite{Cohnfir} and uses
  Ronyai's algorithm for computing common invariant subspaces of a
  collection of matrices over finite fields \cite{Ronyai2}.\\

\item \textbf{Recovering the factors of $M$}~~ This part requires a
  new algorithm but similar to the case of a single polynomial
  \cite{AJ22}. The factor recovery algorithm is based on an algorithm
  for trivializing a matrix product relation of the form $AB=0$, where
  $A\in\F_q\angle{X}^{s\times r}$ is a linear matrix and
  $B\in\F_q\angle{X}^{r\times t}$ is a matrix of polynomials. We
  efficiently compute an invertible matrix $N\in\F_q\angle{X}^{r\times
    r}$ such that $N^{-1}$ is also in $\F_q\angle{X}^{r\times r}$
  (which means $N$ is a unit in the ring $\F_q\angle{X}^{r\times r}$).
  Then $AB=0$ can be recast as $(AN)(N^{-1}B)=0$ which is trivialized
  by $N$ in the following sense: for every index $i\in [r]$ either the
  $i^{th}$ column of $AN$ is zero or the $i^{th}$ row of $N^{-1}B$ is
  zero. Using this algorithm repeatedly we are able to recover the
  factorization of $M$ from the factorization of $L$.
\end{itemize}

\noindent\textbf{Plan of the paper.}~ In Section~\ref{nc-mat-fact-sec}
we present the algorithm for factorization of matrix polynomials over
the noncommutative ring $\F_q\angle{X}$. In
Section~\ref{univ-matf-sec} we present a deterministic polynomial time
algorithm for factorization of matrix polynomials over the commutative
polynomial ring $\Q[x]$. 


\section{Factorization of matrix polynomials over $\F_q\angle{X}$}\label{nc-mat-fact-sec}

In this section we prove the main theorem showing a randomized
polynomial-time algorithm for factorization of full non-unit matrices
over $\F_q\angle{X}$. As explained in the introduction, the algorithm
strategy has three broad steps:
\begin{itemize}
\item[(a)] Higman linearization of the input matrix polynomial $M$
  which produces a linear matrix $L$ which is an associate of $M$. The
  input matrix $M$ is assumed of full noncommutative rank (hence $L$
  will also be of full rank).
\item[(b)] Factorization of the linear matrix $L$ by using Cohn's
  factorization theory to reduce the problem to computation of a
  common invariant subspace for a collection of scalar matrices over
  $\F_q$ which can be solved in randomized polynomial-time using
  Ronyai's algorithm \cite{Ronyai2}.
\item[(c)] The efficient recovery of the factors of $M$ from the
  factors of $L$.
\end{itemize}  

The same strategy was used in \cite{AJ22} for the factorization of
noncommutative polynomials. The algorithms for the first two steps (a)
and (b) for matrix polynomials follow from the results in \cite{AJ22}.
Efficient computation of Higman linearization works for matrix
polynomials as well \cite{GGOW20}.

\begin{theorem}\label{ehigman}{\rm\cite[Proposition A.2]{GGOW20}}
Let $M \in \fR^{n \times n}$ such that $M_{i,j}$ is computed by a
non-commutative arithmetic formula of size at most $s$ and bit
complexity at most $b$. Then, for $\ell=O(s)$, in time $\poly(s,b)$ we
can compute the matrices $P, Q$ and $L$ in $\fR^{(n+\ell)\times(n+\ell)}$
of Higman's linearization such that
\[
\left(
\begin{array}{c|c}
M & 0 \\
\hline
0 & I_{\ell}
\end{array}
\right) ~=~PLQ.
\]
Moreover, the entries of the matrices $P$ and $Q$ as well as $P^{-1}$
and $Q^{-1}$ are given by polynomial-size algebraic branching programs
which can also be obtained in polynomial time.
\end{theorem}

Noncommutative linear matrix factorization is already dealt with in
\cite{AJ22} as mentioned above.

\begin{theorem}{\rm\cite{AJ22}}\label{lfact2}
On input a full and right (or left) monic linear matrix $L= A_0 +
\sum_{i=1}^n A_i x_i$ where $A_i \in \mathbb{F}^{d \times d}$ for
$i\in [n]$, there is a randomized polynomial time $\poly(n,d,\log_2q)$
algorithm to compute a factorization $L=F_1F_2\cdots F_r$, where each
$F_i$ is a linear matrix atom. Furthermore, there is a deterministic
time $\poly(n,d,q)$ algorithm for the problem. 
\end{theorem}

\begin{remark}\label{lin-fact-rem}
  We explain the above theorem statement in detail. The linear matrix
  $L$ is called right monic if the matrix $[A_1 A_2 \cdots A_n]$ has
  full row rank \cite{Cohnfir}.  As explained in \cite{AJ22}, the
  factorization problem for full linear matrices can be reduced to the
  factorization of full and right monic linear matrices. Furthermore,
  the factorization algorithm for linear matrices in \cite{AJ22} is
  based on Ronyai's common invariant subspace algorithm, and that
  reduction requires $A_0$ to be invertible. Transforming $L$ to
  fulfill this property will, in general, requires a blow-up in the
  matrix dimension. This is achieved by first finding a matrix
  substitution $x_i\leftarrow M_i$ such that $L(M_1,M_2,\ldots,M_n)$
  is invertible. This can be computed in deterministic polynomial time
  using the noncommutative rank algorithm of \cite{IQS18} (specifically, see Section 1.2 in \cite{IQS18})\footnote{In
    \cite{AJ22} we used a different result \cite{DM17} for this
    purpose which, in randomized polynomial time, gives such a matrix
    substitution with entries of matrices from possibly some extension
    field of $\F_q$. However, in \cite{IQS18} such a matrix
    substitution is obtained in deterministic polynomial time.
    Moreover, it can be ensured that the entries of the obtained
    matrices reside in $\F_q$ itself even for small $q$, at the cost
    of slightly larger dimensional matrix substitution.},
   where $M_i$ are matrices of dimension
  $r=\poly(s,d)$. Then, the substitution $x_i\leftarrow Y_i + M_i$,
  where each $Y_i$ is $r\times r$ matrix of fresh noncommuting
  variables, results in a blown-up $rd\times rd$ linear matrix $L'$
  whose constant term $A'_0$ is now invertible. The linear matrix
  factorization algorithm in \cite{AJ22} (Theorem~\ref{lfact2}) finds
  a complete factorization of $L'$ and, in polynomial time, recovers
  from it a complete factorization of $L$.

  In summary, we observe that in fact the linear matrix factorization
  algorithm of \cite{AJ22} is really a deterministic polynomial-time
  reduction to the problem of univariate polynomial factorization over
  $\F_q$. Hence, we have a randomized $\poly(n,d,\log_2q)$ time
  algorithm for it and, alternatively, a deterministic time
  $\poly(n,d,q)$ time-bounded algorithm.
  \end{remark}

The new contribution is in step (c) for recovering the factorization
of $M$ from the factorization of the linear matrix $L$. We now
proceed to describe this algorithm.

\subsection{Trivialization of matrix relations}\label{triv-subsec}

For matrices $C\in\fR^{d\times r}$ and $D\in\fR^{r\times s}$ the
condition
\[
CD=0
\]
is called a \emph{matrix relation} \cite{Cohnfir}. A unit
$M\in\fR^{r\times r}$ is said to \emph{trivialize} the matrix relation
$CD=0$ if for every index $i, 1\le i\le r$ either the $i^{th}$ column
of the matrix $CM$ is all zeros or the $i^{th}$ row of the matrix
$M^{-1}D$ is all zeros.

The existence of $M$ is proved using an argument about bases for
finite-dimensional modules in Cohn's book \cite{Cohnfir}. However, a
natural algorithmic problem is the complexity of computing the matrix
unit $M$. For matrices over fields (finite fields or $\Q$) $M$ can be
found in polynomial time using standard linear algebraic
computation. In \cite{AJ22} we give a deterministic polynomial-time
algorithm when $C\in\fR^{d\times r}$ is a linear matrix and
$D\in\fR^{r\times 1}$ is a column vector of polynomials, where the
underlying field is a finite field. The algorithm computes $M$ and its
inverse $M^{-1}$ in $\fR^{r\times r}$ such that their entries are
given by polynomial-size algebraic branching programs.

For the problem of matrix factorization considered in this paper, we
require a generalization of this to the case when $C$ is a linear
matrix and $D$ is a matrix of polynomials. A simple trick allows us to
adapt the algorithm of \cite{AJ22}.


\begin{lemma}\label{triv-lem}
  Let $C\in\fR^{k \times \ell}$ be a linear matrix and $\tilde{D} \in
  \fR^{\ell \times m}$ be a matrix polynomial with entries of
  $\tilde{D}$ are given by algebraic branching programs such that $C
  \tilde{D}=0$.
  Then, in deterministic polynomial time we can compute a invertible
  matrix $N\in\fR^{\ell \times \ell}$ such that
  \begin{itemize}
  \item For $1\le i\le \ell$ either the $i^{th}$ column of $CN$ is all zeros
    or the $i^{th}$ row of $N^{-1}\tilde{D}$ is zero.
  \item Each entry of $N$ is a polynomial of degree at most $\ell^2$ and
    is computed by a polynomial size ABP, and also each entry of
    $N^{-1}$ is computed by a polynomial size ABP.
  \end{itemize}
\end{lemma}

\begin{proof}
Let $\tilde{D}_i$ denotes the $i^{th}$ column of $\tilde{D}$ for $i\in
[m]$. Let $v \in\fR^{\ell \times 1}$ be a column of polynomials
defined as, $v = \sum_{i=1}^m \tilde{D}_i y^i$ where $y \not\in X$ is
a fresh noncommuting variable. We clearly have, $C\tilde{D}=0$ if and
only if $C\tilde{D}_i=0$ for all $i\in [m]$ if and only if $Cv=0$.

To keep the paper self-contained we reproduce the algorithm from
\cite{AJ22}. We describe it as a recursive procedure Trivialize that
takes matrix $C$ and column vector $v$ as parameters and returns a
matrix $N$ as claimed in the statement. The key additional point to
note is that the variable $y$ occurs only in $v$ and not in $C$, and
the construction of $N$ is such that its entries are only polynomials
in variables used in $C$.

\begin{enumerate}
\item[] Procedure Trivialize$(C\in\fR^{k\times \ell},v\in\fR^{\ell\times 1})$
\item If $d=1$ then (since $Cv=0$ iff either $C=0$ or $v=0$ because
  the ring $\fR$ has no zero divisors) \textbf{return} the identity
  matrix.
  \item  If $d>1$ then
  \item write $C=C_0+C_1$, where $C_0$ is a scalar matrix and $C_1$ is the
    degree $1$ homogeneous part of $C$. Let $e$ be the degree of the
    highest degree nonzero monomials in the polynomial vector $v$, and
    let $m$ be a nonzero degree $e$ monomial. Let
    $v(m)\in\F_q^{\ell\times 1}$ denote its (nonzero) coefficient in $v$.
    Then $Cv=0$ implies $C_1 v(m)=0$. Let $T_0\in\F_q^{\ell \times \ell}$ be a
    scalar invertible matrix with first column $v(m)$ obtained by
    completing the basis.
    \begin{enumerate}
    \item If $C_0v(m)=0$ then the first column of $CT_0$ is zero.
    \item Otherwise, $CT_0$ has first column as the nonzero scalar
      vector $Cv(m)=C_0v(m)$. Suppose $i^{th}$ entry of $Cv(m)$ is a
      nonzero scalar $\alpha$. With column operations we can drive the
      $i^{th}$ entry in all other columns of $CT_0$ to zero. Let the
      resulting matrix be $CT_0T_1$ (where the matrix $T_1$ is
      invertible as it is a product of elementary matrices
      corresponding to these column operations, each of which is of
      the form $\col_i\leftarrow (\col_i +
      \col_1\cdot \alpha_0+\sum_i \alpha_i x_i)$). Notice that $CT_0T_1$ is still
      linear.
    \item As $Cv=(CT_0T_1)(T_1^{-1}T_0^{-1}v)$, and in the $i^{th}$
      row of $CT_0T_1$ the only nonzero entry is $\alpha$ which is in
      its first column, we have that the first entry of
      $T_1^{-1}T_0^{-1}v$ is zero.
    \end{enumerate}
  \item Let $C'\in\fR^{k \times (\ell-1)}$ obtained by dropping the first
    column of $CT_0T_1$. Let $v'\in\fR^{(\ell-1)\times 1}$ be obtained by
    dropping the first entry of $T_1^{-1}T_0^{-1}v$. Note that $C'$
    is still linear.
  \item Recursively call Trivialize$(C'\in\fR^{k\times
    (\ell-1)},v'\in\fR^{(\ell-1)\times 1})$.  and let the matrix returned by
    the call be $T_2\in\fR^{(\ell-1)\times (\ell-1)}$.
  \item Putting it together, return the matrix $T_0T_1(I_1\oplus T_2)$.
  \end{enumerate}

Now, since $v = \sum_{i=1}^m \tilde{D}_i y^i$, for all $j\in [\ell]$,
we have $j^{th}$ entry of $v$ is equal to $0$ if and only if $j^{th}$
entry of all the columns $\tilde{D}_i$ is equal to zero for $i\in
[m]$. Consequently we get, $j^{th}$ entry of $v$ is equal to $0$ iff
$j^{th}$ row of $\tilde{D}$ is all zero for all $j\in [\ell]$. So the
matrix $N$ satisfies the required property, namely for $1\le i\le
\ell$ either the $i^{th}$ column of $CN$ is all zeros or the $i^{th}$
row of $N^{-1}\tilde{D}$ is all zero.
 
To complete the proof, we note that a highest degree monomial $m$ of
the vector $v$ such that its (scalar) coefficient vector $v(m)\ne 0$
is easy to compute in deterministic polynomial time when each
polynomial $v_i$ is given by a noncommutative algebraic branching
program: we can use the PIT algorithm of Raz and Shpilka \cite{RS05}
to find the coefficient of $m$ in each $v_i$ to obtain the vector
$v(m)$. Now, for the recursive call we require $C'$ to be also a
linear matrix and each entry of the new polynomial vector $v'$ to have
a small ABP. The matrix $C'$ is linear because $CT_0T_1$ is a linear
matrix: because $CT_0$ is linear and its first column is scalar, and
each column operation performed by $T_1$ is scaling the first column
of $CT_0$ by a linear form and subtracting from another column of
$CT_0$. Regarding the polynomial vector $v'$, each entry of it has a
small ABP because $T_0^{-1}$ is scalar and the entries of the matrix
$T_1^{-1}$ have ABPs of polynomial size because $T_1$ (and hence also
$T_1^{-1}$) is a product of units which correspond to elementary
column operations. Finally, we note that $T_1$ is a product of at most
$\ell-1$ linear matrices each corresponding to a column operation, and
$N$ is an iterated product of $\ell$ such matrices. Hence, each entry
of $N$ as well as $N^{-1}$ is a polynomial of degree at most $\ell^2$
and is computable by a small ABP.
\end{proof}

\subsection{Factor extraction algorithm}\label{extr-subsec}

We first recap the overall algorithm to see where the factor
extraction algorithm will fit in. For an input $M\in\fR^{d\times d}$,
which is a full matrix whose entries are computed by noncommutative
arithmetic formulas, we apply Higman linearization \cite{GGOW20} to
compute in polynomial time the following
\begin{equation}\label{eq5}
  \left(
\begin{array}{cc}
M & 0 \\
0 & I
\end{array}
\right)
= PLQ,
\end{equation}

where $L$ is a full linear matrix, $P$ is an upper triangular matrix
will all ones diagonal, $Q$ is a lower triangular matrix with all ones
diagonal and the entries of $P$ and $Q$ are computed by noncommutative
ABPs.

Next, we apply the linear matrix factorization algorithm of
\cite{AJ22} to $L$ and compute a complete factorization as
\[
L = L_1L_2\cdots L_t,
\]
where each $L_i$ is a linear matrix atom.  The factor extraction
algorithm will find a complete factorization of the matrix
$M=M_1M_2\cdots M_t$ as a product of $t$ matrix atoms. It is based on
the following lemma which will allow us to recover the factors one by
one.

\begin{lemma}[Factor Extraction Lemma]\label{extract}
  Let $M \in\fR^{m \times m}$ be a matrix polynomial and $V \in\fR^{k
    \times k}$ be a unit with
  \begin{equation}\label{eq5}
  \left(
\begin{array}{cc}
M & U \\
0 & V
\end{array}
\right)
= P CD,
\end{equation}
such that
\begin{itemize}
\item $U\in \fR^{m \times k}$ is a matrix polynomial, $C\in \fR^{\ell
  \times \ell}$ is a full linear matrix that is a non-unit, $P\in
  \fR^{\ell \times \ell}$ is upper triangular with all $1$'s diagonal,
  and $D\in\fR^{ \ell \times \ell}$ is a full non-unit matrix which is
  also an atom, where $\ell = m+k$.
\item The entries of $M, U, V, P, D$ are all given as input by
  algebraic branching programs and the linear matrix $C$ is given
  explicitly.
\end{itemize}
Then we can compute in deterministic polynomial time a nontrivial
factorization $M =G \cdot H$ of the matrix $M$ where both $G$ and $H$
are full non-unit matrices and, moreover, the matrix $H$ is an atom.
\end{lemma}

\begin{proof}
Let
\[C=\left(
\begin{array}{cc}
c_1 & c_3 \\
c_2 & c_4
\end{array}
\right) \text{ and }
D = 
  \left(
\begin{array}{cc}
d_1 & d_3\\
d_2 & d_4
\end{array}
\right),
\]
written as $2\times 2$ block matrices where $c_1$, $d_1$ are
$m \times m$ blocks and $c_4$ and $d_4$ are
$k \times k$ blocks.  By dropping the first $m$ rows of the matrix in the
left hand side of Equation~\ref{eq5} and the first $m$ row of $P$ we
get
\[
  (0~V) = (0~P') C D,
\]
where $P'$ is also an upper triangular matrix with all $1$'s diagonal.
Equating the block consisting of the first $m$ columns on both sides we have
\begin{eqnarray*}
  0 & =&  (0~P')\left(
\begin{array}{cc}
c_1 & c_3 \\
c_2 & c_4
\end{array}
\right)\left(\begin{array}{c} d_1\\ d_2 \end{array} \right),
  \text{ which implies that}\\
  0 & = & P'(c_2~c_4)\left(\begin{array}{c} d_1\\ d_2 \end{array} \right),
  \text{ and hence}\\
  0 & = & (c_2~c_4)\left(\begin{array}{c} d_1\\ d_2 \end{array} \right),
  \text{ since $P'$ is invertible.}
\end{eqnarray*}


Since $(c_2~c_4)\in\fR^{k \times \ell}$ is a matrix with linear
entries and $\left(\begin{array}{c} d_1\\ d_2 \end{array}
\right)\in\fR^{\ell \times m}$ is a matrix of polynomials which are
given by ABPs as input, this is a matrix relation to which we can
apply the trivializing algorithm of Lemma~\ref{triv-lem}. The
trivializing algorithm computes a unit $N$ whose entries are all given
by ABPs such that for $1\le i\le \ell$, either the $i^{th}$ column of
$(c'_2~c'_4)=(c_2~c_4)N$ is all zero or the $i^{th}$ row of
$\left(\begin{array}{c} d_1'\\ d_2' \end{array}
\right)=N^{-1}\left(\begin{array}{c} d_1\\ d_2 \end{array} \right)$ is
all zero.



Since $D$ is a full matrix, the matrix $N^{-1}D$ is also full which
implies $\left(\begin{array}{c} d_1'\\ d_2' \end{array} \right)$ has
at least $m$ non-zero rows as stated in the Claim below. 

\begin{claim}
The $\ell \times m$ matrix $\left(\begin{array}{c}
  d_1'\\ d_2' \end{array} \right)$ has at least $m$ non-zero rows.
\end{claim}

As at least $m$ rows of the matrix $\left(\begin{array}{c}
  d_1'\\ d_2' \end{array} \right)$ are nonzero, it follows that at
least $m$ columns of $(c'_2~c'_4)$ are all zeros because the matrix
$N$ trivializes the relation (that is, for every $i\in [\ell]$ we have
either $i^{th}$ column of $(c'_2~c'_4)$ is all zero or the $i^{th}$
row of $\left(\begin{array}{c} d_1'\\ d_2' \end{array} \right)$ is all
zeros). Hence, there exists a permutation matrix $\Pi$ such that the
first $m$ columns of $(c'_2~c'_4)\Pi$ are all zeros \emph{and} the
first $m$ rows of $\Pi^{-1} \left(\begin{array}{c}
  d_1'\\ d_2' \end{array} \right)$ are all nonzero.


Consider the matrices $C''=CN\Pi=\left(
\begin{array}{cc}
c''_1 & c''_3 \\
c_2'' & c''_4
\end{array}
\right)
$ and
$D''=\Pi^{-1}N^{-1}D=\left(
\begin{array}{cc}
d''_1 & d''_3 \\
d_2'' & d''_4
\end{array}
\right)
$. Let
\[
P =\left(
\begin{array}{cc}
P_1 & * \\
0 & P_2
\end{array}
\right),
\]
where both $P_1$ and $P_2$ are upper triangular matrices with all ones
diagonal.  Then, we have

\[P^{-1}
\left(
\begin{array}{cc}
M & U \\
0 & V
\end{array}
\right)
=
\left(
\begin{array}{cc}
P_1^{-1}M & * \\
0 & P_2^{-1}V
\end{array}
\right)
=
\left(
\begin{array}{cc}
c''_1 & c''_3 \\
c_2'' & c''_4
\end{array}
\right)
  \left(
\begin{array}{cc}
d''_1 & d''_3\\
d_2'' & d''_4
\end{array}
\right),
\]
where $c_2''$ is all zero block of size $k \times m$ and $d_1''$ is
matrix of size $m \times m$ such that all rows of $d_1''$ are
non-zero. By looking at $(2,2)^{th}$ block in the above equation, we
can see that $c_4''$ and $d_4''$ are units as $P_2^{-1}V$ is a
unit. Now observing $(2,1)^{th}$ matrix block in the above equation,
we get $d_2''$ is all zero as $c_4''$ is a unit.  Clearly, we have
$P_1^{-1}M= c_1'' \cdot d_1 ''$. Now, since $C$ and $D$ are non-units
(by assumption), the matrices $C''$ and $D''$ are also
non-units. Therefore, $c''_1$ is not a unit for otherwise $C''$ would
be a unit. Similarly, $d''_1$ is not a unit.  It follows that
$M=P_1c''_1d''_1$ is a nontrivial factorization of $M$, noting that
$P_1$ is a unit (being upper triangular with all ones diagonal).

Furthermore, since $D$ is an atom and $D''=\Pi N^{-1}D$ where both
$\Pi$ and $N$ are units in $\fR^{d\times d}$ it follows that $D''$ is
also an atom. As $D'' = \left(
\begin{array}{cc}
d''_1 & d''_3\\
0 & d''_4
\end{array}
\right)$ and $d_4''$ is invertible, we get
\[
\left(
\begin{array}{cc}
I_m & 0\\
0 & (d''_4)^{-1}
\end{array}
\right) \cdot D'' = \left(
\begin{array}{cc}
d''_1 & d''_3\\
0 & I_{k}
\end{array}
\right).
\]
Now applying suitable row operations to the matrix $(I_m \oplus
(d''_4)^{-1})D''$ we can drive $d_3''$ to zero. So we have $W(I_m
\oplus (d''_4)^{-1})D''=(d_1'' \oplus I_k)$ for a unit $W$. Hence
$d_1''$ is an associate of $D''$ and therefore $d_1''$ is an atom
because $D''$ is an atom.

Thus, we conclude that the above algorithm computes a nontrivial
factorization of $M$ as
\[
M= G\cdot H
\]
where $G=P_1^{-1}c_1''$ is a full non-unit matrix, and $H=d_1''$ is an
atom, and $P_1$ is a unit.
\end{proof}

\subsection{The Factorization Algorithm}

We now put everything together and describe the factorization algorithm.

\begin{theorem}\label{matfactthm}  
  Let $\fR=\F_q\angle{X}$ and let $M\in \fR^{d\times d}$ be a matrix
  of noncommutative polynomials where each of its entries $M_{ij}$ is
  given by an arithmetic formula of size at most $s$ as input
  instance. Then there is a $\poly(s, \log q)$ time randomized
  algorithm that outputs a complete factorization of $M$ as a product
  $M=M_1M_2\cdots M_r$ such that each matrix factor $M_i$ is an atom,
  and the entries of the matrix factors are output as algebraic
  branching programs. Moreover, there is also a deterministic
  $\poly(s, q)$ time deterministic algorithm for the problem.
\end{theorem} 

\begin{proof}
  Given $M\in\fR^{d\times d}$ as input, we apply Higman linearization
  followed by the linear matrix factorization algorithm stated in
  Theorem~\ref{lfact2} (see \cite{AJ22} for details) to obtain the
  factorization
\[
M\oplus I_s=PF_1F_2 \ldots F_rU
\]
where each linear matrix $F_i$ is an atom, the matrix $P$ is upper
triangular with all $1$'s diagonal, and the matrix $U$ is a
unit. Moreover, the entries of $P$ and $U$ are given by algebraic
branching programs.

We will now apply Lemma~\ref{extract} to extract the factors of $M$
(one by one from the right).

  For the first step, let $C=F_1F_2\cdots F_{r-1}$ and $D=F_rU$ in
  Lemma~\ref{extract}. The proof of Lemma~\ref{extract} yields the
  matrix $N_r=N\Pi$ such that both matrices $C''=PF_1F_2\cdots
  F_{r-1}N_r$ and $D''=N_r^{-1}F_rU$ has the first $d$ column all
  zeros except the top left $d\times d$ block of entries $c''_1$ and
  $d''_1$ which yields the nontrivial factorization $M=c''_1d''_1$,
  where $d''_1=M_r$ is an atom. Renaming $c''_1$ as $G_r$ we have from
  the structure of $C''$:
  \[
  \left( \begin{array}{cc} G_r & * \\ 0 & V_r \end{array} \right) =
  P(F_1F_2\cdots F_{r-2}) (F_{r-1}N_r).
  \]
Setting $C= F_1F_2\cdots F_{r-2}$ and $D=F_{r-1}N_r$ in
Lemma~\ref{extract} we can compute the matrix $N_{r-1}$ using which we
will obtain the next factorization $G_r=G_{r-1}M_{r-1}$, where
$M_{r-1}$ is an atom by Lemma~\ref{extract}. Note that
Lemma~\ref{extract} is applicable as all conditions are met by the
matrices in the above equation (note that the matrix $V_r$ will be a
unit).

Continuing thus, at the $i^{th}$ stage we will have
$M=G_{r-i+1}M_{r-i+1}M_{r-i+2}\cdots M_r$ after obtaining the
rightmost $i$ irreducible factors by the above process. At this stage
we will have
  \[
  \left( \begin{array}{cc} G_{r-i+1} & * \\ 0 & V_{r-i+1} \end{array}
  \right) = P(F_1F_2\cdots F_{r-i-1}) (F_{r-i}N_{r-i+1}),
  \]
  where $V_{r-i+1}$ is a unit and all other conditions are satisfied
  for application of Lemma~\ref{extract}.

  Thus, after $r$ stages we will obtain the complete factorization of the input
  matrix $M$ as
\[  
M=M_1M_2\cdots M_r,
\]
where each factor $M_i$ is an atom.\\

\subsubsection*{Running Time Analysis}

The Higman Linearization of $M$ is computed in deterministic
polynomial time. For the resulting linear matrix $L=A_0+\sum_{i=1}^n
A_i x_i$, as explained in Remark~\ref{lin-fact-rem}, its factorization
as a product of linear matrix atoms can be computed in randomized
$\poly(s,n,d,\log_2q)$ time as well as in deterministic
$\poly(s,n,d,q)$ time.

For the rest of the running time, it suffices to note that the matrix
$N$ computed in Lemma~\ref{extract} is a product of degree at most
$d^2$ many linear matrices (corresponding to the column
operations). Thus, at the $i^{th}$ of the above iteration, the sizes
of the ABPs for the entries of $N_{r-i+1}$ are independent of the
stages. Hence, the overall randomized algorithm has running time
$\poly(s,n,d,\log_2q)$. The deterministic factorization algorithm
has running time $\poly(s,n,d,q)$.
\end{proof}

\section{Factorization of matrices over $\Q[x]$}\label{univ-matf-sec}

In this section we describe a deterministic polynomial-time algorithm
for the factorization of full non-unit matrices with univariate
polynomial entries over rationals (i.e., the matrix entries are from
$\Q[x]$). As there is only one variable $x$, the noncommutative ring
$\F\angle{x}$ coincides with the commutative ring $\F[x]$. Thus, over
a finite field $\F=\F_q$ we note that the algorithm of the previous
section also solves matrix factorization over $\F_q[x]$ as a special
case.

Our technique is essentially the same: we first transform the input
matrix $M$ into a linear matrix $L$ by Higman linearization (see
Theorem~\ref{ehigman}).  We obtain
\[
\left(
\begin{array}{c|c}
M & 0 \\
\hline
0 & I_{\ell}
\end{array}
\right) ~=~PLQ.
\]
where $P, Q$ and $L$ are matrices with entries from $\Q[x]$.

As mentioned in the introduction, the problem of factorization of
multivariate noncommutative polynomials over $\Q$ (i.e., over the free
noncommutative ring $\Q\angle{X}$) is not amenable to our approach
\cite{AJ22} because the problem of computing a nontrivial common
invariant subspace for a set of matrices over $\Q$ seems intractable
in general \cite{Ron87}. However, in the univariate case we need to
compute a nontrivial invariant subspace for a single rational matrix
which can be done efficiently using basic linear algebra. This gives
us a deterministic polynomial-time algorithm for factorizing matrix
polynomials over $\Q[x]$.

In the first subsection below we will present an efficient
trivializing algorithm for the matrix relation $C D=0$ where
$C\in\Q[x]^{d\times r}$ is a univariate linear matrix and
$D\in\Q[x]^{d\times s}$ is a matrix of univariate polynomials over
rationals. In the next subsection we will present the algorithm for
factorization of linear matrices $L$ in one variable $x$ over $\Q$.

\subsection{Trivializing matrix relations algorithm over $\Q$}

Let $C\in\Q[x]^{d\times r}$ and $U\in\Q[x]^{r\times s}$ be given as
input such that $CU=0$, where $C$ is a linear matrix. We describe a
polynomial-time algorithm for computing an invertible matrix $N$
such that:
\begin{enumerate}
\item The matrix $N\in\Q[x]^{r\times r}$ is a unit: $\det N$ is a
  nonzero scalar (equivalently $N^{-1}$ is a matrix with polynomial
  entries).
\item The matrix relation $(CN)(N^{-1}U)=0$ is trivialized: for
  each $i\in[r]$ either the $i^{th}$ column of $CN$ is all zeros
  or the $i^{th}$ row of $N^{-1}U$ is all zeros.
\end{enumerate}  

We note that the algorithm we have already described in
Section~\ref{triv-subsec} solves the problem over $\F_q$ in the
multivariate case. That algorithm performs a polynomial number of
arithmetic operations over $\F_q$. However, working over $\Q$ we need
to additionally control the binary encoding lengths of the matrix
coefficients that will result from the repeated column operations. As
such it is not clear to us that the above-mentioned algorithm over
$\F_q$ has this additional property when we use it for $\Q$. However,
we present direct a polynomial-time algorithm for this problem over
$\Q$ in the \emph{univariate case}. 

Let $C= C' +C'' x\in\Q[x]^{d\times r}$. The heart of the
trivialization algorithm is to first efficiently transform $C$ into a
suitable normal form (which we refer to as \emph{T-normal form}). It
turns out that once we have the matrix $C$ in T-normal form it is easy
to compute a trivializing matrix $N$ as required. We will first define
the T-normal form and show that if the linear matrix $C$ is in
T-normal form, the trivializing matrix $N$ can be computed in
polynomial time (taking into account the binary encoding lengths of
all integers involved). Then we will give a polynomial-time algorithm
to transform $C$ into T-normal form.

Let $D \in \Q[x]^{d \times k}$ be a linear matrix and let $D_i$ denote
the $i^{th}$ column of $D$, $1\le i\le k$. Writing $D=D'+D''x$, where
$D',D''\in \Q{d\times k}$, for $1\le i \le k$ we have
\[
D_i =D_i'+D_i''x,
\]
where $D_i'$ and $D_i''$ are the $i^{th}$ columns of $D'$ and $D''$,
respectively. The $i^{th}$ column $D_i$ of $D$ is a \emph{scalar
  column} if all the entries of $D_i$ are scalars (i.e. $D_i''=0$)
otherwise it is a \emph{non-scalar column}.\smallskip

\noindent\textbf{Encoding sizes.}~~ For an integer $a$, the encoding
size of $a$ (denoted as $\bit(a)$) is the number of bits required to
express $a$ in binary. For a rational number $r=\frac{a}{b}, b\neq 0$,
the encoding size of $r$ is $\bit(r)= \max
(\bit(a),\bit(b))$. Extending this further, for a univariate
polynomial $f = a_0 +a_1 x + \ldots + a_t x^t \in \Q[x]$, we define
the binary encoding size of $f$ as $\bit(f)= t\cdot \max_{i=0}^t
\bit(a_i)$. For an univariate polynomial matrix $C \in \Q[x]^{d \times
  k}$, the encoding size of $C$ is, $\bit(C)= kd\cdot \max_{1\leq i
  \leq d, 1\leq j \leq k} \bit(C_{i,j})$ where $C_{i,j}$ is the
$(i,j)^{th}$ entry of $C$.

Let $C$ be a $d \times k$ matrix. For index sets $I\subseteq [d], J
\subseteq [k]$, let $C(I,J)$ denote the submatrix of $C$ with rows
from $I$ and columns from $J$.

\begin{definition}{\bf [T-normal form]}
  A linear matrix $D=[0~A~B]\in \Q[x]^{d \times r}$ is said to be in
  \emph{T-normal form} if its columns are partitioned into the three
  parts, as indicated, such that:
  \begin{itemize}
\item The first set of columns of $D$ is all zeros, followed by the
  second set $A$ of scalar columns, and then the third set $B=B'+B''x$
  consists of non-scalar columns.
\item The matrix $[A~B'']$ is of full column rank.
  \end{itemize}
\end{definition}  

Before we proceed we note that, analogous to
Section~\ref{triv-subsec}, it is convenient to transform the given
matrix relation $CU=0$ into another relation $Cu=0$ where $u$ is a
column vector whose entries are bivariate polynomials in $\Q[x,y]$,
where $y$ is a fresh commuting variable.

For a matrix $U\in\Q[x]^{r\times s}$, we define the column vector
$u\in\Q[x,y]^{r\times 1}$ as
\[
u = \sum_{j=1}^s U_j y^j,
\]
where $U_j, 1\le j\le s$ are the $s$ columns of the matrix $U$. We
note that $CU=0$ if and only if $Cu=0$. We also have the following.

\begin{lemma}
  A matrix $N\in\Q[x]^{r\times r}$ trivializes the relation
  $(CN)(N^{-1}U)=0$ if and only if it trivializes the relation
  $(CN)(N^{-1}u)=0$.
\end{lemma}

We first show that the relation $Du=0$ is easy to trivialize for a
linear matrix $D \in \Q[x]^{d \times r}$ which is in T-Normal form.
  
\begin{lemma}
  Let $D \in \Q[x]^{d \times r}$ be a linear matrix in T-normal form
  and $u\in\Q[x,y]^r$ be a column vector of polynomials given as input
  such that the matrix relation $Du= 0$ holds. Then there is a
  polynomial ($\poly(d,r,\bit(D),\bit(u))$) time deterministic
  algorithm to compute a full rank matrix $N \in \Q[x]^{r \times r}$
  such that $\bit(N) \leq \poly(d,r) \cdot \bit(D)$, the matrix
  $N^{-1} \in \Q[x]^{r \times r}$ (i.e., $N$ is a unit in the ring
  $\Q[x]^{r\times r}$), and the relation $(DN)(N^{-1}u)=0$ is
  trivialized.
\end{lemma}

\begin{proof}
  Let $D=[0~A~B]$ and $J_1, J_2$ and $J_3$ be the column indices of
  the three parts: $0$, $A$ and $B$ witnessing that $D$ is in T-normal
  form. The columns of $A$ are linearly independent. So there is a
  subset of row indices $I\subseteq [k]$ such that $D[I,J_2]$ is a
  full rank square submatrix of $A$. Hence, each column $D(I,j), j\in
  J_3$ of the corresponding submatrix of $B$ can be expressed as
  \[
  D(I,j) = \sum_{i \in J_2} (a_{ji} + b_{ji} x)D(I,i), ~~ j\in J_3
\]
  where $a_{ji}$ and $b_{ji}$ are rational numbers. We can compute
  these numbers $a_{ji}$ and $b_{ji}$ by Cramer's rule. Hence,
  $\bit(a_{ji}),\bit(b_{ji}) \leq \poly(k,n) \bit(D)$ for all column
  indices $j\in J_3$.

Now let $N$ be $d\times d$ column transformation matrix which
implements the column operations $D_j = \sum_{i \in J_2} (a_{ji} +
b_{ji} x)D_i$ for all column indices $j\in J_3$. We note the
following.
\begin{claim}
$N$ is an $r\times r$ upper triangular matrix with all diagonal
  entries $1$ and
  \[
  N_{i,j}= -(a_{j,i} + b_{j,i} x) ~~~ \text{ for }~ i \in J_2~ \text{ and }~ j\in J_3.
  \]
Furthermore, $N^{-1} \in \Q[x]^{r \times r}$ and can be
efficiently computed.
\end{claim}
After performing these column operations we have the matrix
$E=DN=[0~A~C]$, where $C$ is zero on all the rows indexed by $I$. Let
$N^{-1}u = w = [w_1~w_2~\ldots~w_r]^T$. The row indices of $w$ can be
correspondingly partitioned into $[r] = J_1\sqcup J_2\sqcup J_3$. We
denote the corresponding subvectors of $w$ by $w_{J_1}$, $w_{J_2}$ and
$w_{J_3}$. Since $E(I,J_3) = 0$ and $E(I,J_1)=0$ it follows that
\[
E(I,J_2)w_{J_2}=0.
\]
As $E(I,J_2)= D(I,J_2)$ is an invertible scalar matrix it follows that
the subvector $w_{J_2}=0$. Therefore, we have
\[
Ew= Cw_{J_3} = 0.
\]
Now, consider the submatrix $C=C'+C''x$ of $E$. Since the matrix
$D=[0~A~B]$ is in T-normal form, the matrix $[A~B'']$ has full column
rank where $B=B'+B''x$ and both $B'$ and $B''$ are scalar matrices.
Now, since $C$ is obtained from $A$ and $B$ by the column operations
defined by $N$ notice that the matrix $[A~C'']$ is also of full column
rank because the columns of $C''$ are
\[
E''_j = D''_j +\sum_{i\in J_2}b_{ji} D_i, ~~\text{ for all } j\in J_3.
\]

Hence $C''$ is full column rank. Let $K\subset [d]$ be row indices
such that $D''[K,J_3]$ is an invertible submatrix of $C''$. Then the
submatrix $D[K,J_3]$ of the linear matrix $C$ is also invertible in
the field of fractions $\Q(x)$. Therefore, $Cw_{J_3}=0$ forces
$w_{J_3}=0$.

Putting it together, we have shown that the matrix $N$ trivializes the
relation $Dv=(DN)(N^{-1}u)=0$. This completes the proof.
\end{proof}

Now we describe a polynomial-time algorithm that transforms the input
linear matrix $C=C'+C''x$ into a linear matrix $D$ in T-normal form.

\begin{lemma}\label{tnorm-form}
  Given as input a linear matrix $C=C'+C''x\in\Q[x]^{k\times d}$ in
  deterministic time $\poly(k,d,\bit(C))$ we can compute a matrix
  $M\in\Q[x]^{d\times d}$ such that
  \begin{itemize}
  \item $D=CM$ is in T-normal form.
  \item The $d\times d$ matrix $M$, which is a product of elementary
    column operation matrices, is a unit in $\Q[x]^{d\times d}$ (i.e.,
    it has nonzero scalar determinant).
 \end{itemize}
\end{lemma}

\begin{proof}
  We describe the algorithm along with the correctness of each step,
  side by side.

  \begin{enumerate}
  \item[] Input $C=C'+C''x\in\Q[x]^{k\times d}$.
      \item By permuting the columns of $C$ write it as $[0~A~B]$,
      consisting of a block of $0$ columns followed by a block of
      scalar columns $A$ and then the columns containing the linear
      submatrix $B$.  
    \item By performing column operations on $A$ we can drive all
      linearly dependent columns to zero and move such columns to the
      left. Thus the block of columns $A$ can be assumed to be
      linearly independent.
    \item Let $B=B'+B''x$, with $B'$ and $B''$ scalar.
     \begin{enumerate}
      \item \textbf{while} the matrix $[A~B'']$ is not full column rank \textbf{do}
    \item Let $\{A_i\mid i\in I\}\cup \{B''_j\mid j\in J\}$ be a
      dependent set of columns. Then for some $j_0\in J$ there are
      scalars $\alpha_i,\beta_j\in Q$ such that $B''_{j_0}x =\sum_{j\in
        J\setminus\{j_0\}} \beta_jB''_jx + \sum_{i\in I}\alpha_i A_ix$.
    \item Applying the corresponding column operations we can drive
      $B''_{j_0}$ to zero. Note that during this process, the scalar part of $B_{j_0}$ will also get updated as $B'_{j_0} \leftarrow B'_{j_0}+ \sum_{j\in
        J\setminus\{j_0\}} \beta_jB'_j$.
    \item $A:=A\cup \{B'_{j_0}\}$ and $B'':= B''\setminus \{B''_{j_0}\}$. 
    \item If $B'_{j_0}$ is linearly dependent on $A$ we can drive it to zero. 
    \item \textbf{end-while}
    \end{enumerate}  
    \end{enumerate}
  In order to see the correctness, notice that each time the while
  loop executes the number of columns in $B''$ decreases and the
  number of columns in the submatrix $[0~A]$ increases: if $B'_{j_0}$
  is linearly independent of $A$ then it is included in $A$ and the
  number of columns of $A$ (all linearly independent) increases or we
  can drive $B'_{j_0}$ to zero using columns operations with the
  columns of $A$. Therefore, the number of times the while loop
  executes is bounded by $d$. Hence the overall number of arithmetic
  operations performed is also bounded by $\poly(k,d)$. Now we analyze
  the growth of the coefficients of the matrices $A$ and $B$ as the
  algorithm iterates. Note that whenever we express certain column as
  linear combination of some other columns, using Cramer's rule we can
  polynomially bound all the coefficients involved in the linear
  combination. Now, the
  only step in which a column changes and is used again is when the
  column $B'_{j_0}$ gets modified in the process of driving the column
  $B''_{j_0}$ to zero, and then the modified column $B'_{j_0}$ is used
  again as part of the set $A$.  Crucially, we note that the columns
  of $A$ do not cause the change in coefficients of $B'_{j_0}$. It is
  only modified by the coefficients $\beta_j, j\in J\setminus\{j_0\}$
  because in the linear combination the columns $A_i, i\in I$ are
  multiplied by $\alpha_i x$. Thus, it follows the encoding sizes of
  all rational numbers involved in the matrix $[0~A~B]$ at any stage
  of the computation remains polynomially bounded in
  $\bit(C)$. Finally, we note that the matrix $M$ is a product of
  $\poly(d)$ many elementary matrices, corresponding to the elementary
  column operations. Since the entries of $[0~A~B]$ has polynomially
  bounded encoding size in all stages of the computation, the rational
  entries in each such elementary matrix also has encoding size
  polynomially bounded in $\bit(C)$. This completes the proof of the
  lemma.
\end{proof}

Putting it together we have show the following.

  \begin{theorem}\label{triv-lem-rat}
    Given the matrix product relation $C U=0$, where
    $C\in\Q[x]^{d\times r}$ is a linear matrix and $U\in\Q[x]^{r\times
      s}$ is a matrix of polynomials, in deterministic polynomial time
    (in bit complexity) we can compute an invertible matrix $N\in
    \Q[x]^{d\times d}$ such that its inverse $N^{-1}\in \Q[x]^{d\times
      d}$ such that the matrix product $(CN)(N^{-1} U)=0$ trivializes
    the relation $CU=0$.
\end{theorem}   

\subsection{Univariate linear matrix factorization over $\Q$}

The goal of this subsection is a deterministic polynomial-time
algorithm that takes as input a full rank linear matrix $L=A_0 +
A_1x\in\Q[x]^{d\times d}$ and computes a complete factorization of
$L$. We will require two conditions on $L$ before we proceed with the
algorithm.

\begin{definition}\label{monic-def}
  A linear matrix $L=A_0+A_1x$ is called \emph{monic} if the matrix
  $A_1$ is invertible.\footnote{The notions of right and left monic,
   defined in the multivariate setting \cite{Cohnfir,AJ22}, coincide in the
    univariate case which we refer to as simply monic here.}
\end{definition}

\begin{lemma}\label{monic-lem}
  Given a full linear matrix $L=A_0+A_1x\in\Q[x]^{d\times d}$ that is
  not monic we can compute, in deterministic polynomial (in $d$ and $\bit(C)$) time, units
  $U, U'\in\Q[x]^{d\times d}$ and scalar invertible matrices
  $S, S'\in\Q^{d\times d}$ such that
  
  \begin{enumerate}
  \item 
  \[
  ULS =  
\left(
\begin{array}{c|c}
W & 0 \\
\hline
0 & I_{\ell}
\end{array}
\right),
\]
where $W$ is a full monic linear matrix and $\ell>0$. 

\item 
  \[
  S'LU' =  
\left(
\begin{array}{c|c}
W' & 0 \\
\hline
0 & I_{\ell'}
\end{array}
\right),
\]
where $W'$ is a full monic linear matrix and $\ell'>0$. 

\end{enumerate}

\end{lemma}

\begin{proof}
We will prove only the first part, the second part follows symmetrically. 
  We first compute the T-normal form of the transpose matrix
  $L^T=A_0^T +A_1^T x$ by applying the algorithm of
  Lemma~\ref{tnorm-form}. This yields the T-normal form
  \[
  L^T M = [A~B]
  \]
  where the matrix $M\in\Q[x]^{d\times d}$ is a unit: in the T-normal
  form notice that there are no zero columns as $L$ is full rank, and
  the scalar matrix $[A~B'']$ is full rank where $B=B'+B''x$. Let
  $A\in\Q^{d\times e}$. We note that $d>e>0$ as $L$ is a non-unit but
  not monic. We apply the following sets of row/column operations
  on the matrix $[A~B]$:
  \begin{itemize}
  \item Swap the columns of $[A~B]$ to get $[B~A]$.
  \item Since $A$ is a $d\times e$ matrix of rank $e$, we can permute
    the rows and transform $[B~A]$ to $\left(
\begin{array}{c|c}
\hat{B}_1 & \hat{A}_1 \\
\hline
\hat{B}_2 & \hat{A}_2
\end{array}
\right)$ such that the $e\times e$ scalar matrix $\hat{A}_2$ is full rank.

\item Using the invertible $e\times e$ scalar submatrix $\hat{A}_2$ we
  can perform column operations that drives the submatrix $\hat{B}_2$
  to zero to obtain
  $\left(
\begin{array}{c|c}
\hat{B}_3 & \hat{A}_1 \\ \hline 0 & \hat{A}_2
\end{array}
\right)$. Notice that these column operations will be realized by
post-multiplication with a matrix unit $N\in\Q[x]^{d\times d}$ whose
entries are linear in $x$. Moreover, writing
$\hat{B}_3=\hat{B}'_3+\hat{B}''_3x$, we note that the matrix
$\left(
\begin{array}{c|c}
\hat{B}''_3 & \hat{A}_1 \\ \hline 0 & \hat{A}_2
\end{array}
\right)$ has full column rank as $[B''~A]$ has full column rank.
\item Next, with scalar row operations we can use $\hat{A}_2$ to
  drive $\hat{A}_1$ to zero to obtain
$\left(
\begin{array}{c|c}
\hat{B}_3 & 0 \\ \hline 0 & \hat{A}_2
\end{array}
\right)$.
\item Finally, with scalar row and column operations we can drive
  $\hat{A}_2$ to the identity matrix $I_e$ to obtain
 $\left(
\begin{array}{c|c}
\hat{B}_3 & 0 \\ \hline 0 & I_e
\end{array}
\right)$.
\end{itemize}
  Putting it together, we have
\[
S_1L^TM S_2N S_3 =
\left(
\begin{array}{c|c}
\hat{B}_3 & 0 \\ \hline 0 & I_e
\end{array}
\right),
\]
where $S_1, S_2,S_3$ are invertible scalar matrices, and $M$ and $N$
are matrix units. Since $U^T=M S_2N S_3 $ is also a matrix unit, by
again taking transpose we obtain the required
\[
ULS = 
\left(
\begin{array}{c|c}
W & 0 \\ \hline 0 & I_e
\end{array}
\right)
\]
where $W=\hat{B}_3^T$ and $S=S_1^T$. To see that $W$ is monic it
suffices to note that the transformation $N$ is essentially equivalent
to performing column operations on the full rank scalar matrix $[A
 B'']$ which cannot lower the column rank of the resulting matrix. This completes the proof of the part one, for second part, we start with the linear matrix $L$ itself, instead of $L^T$ and carry out all the steps symmetrically.
 
\end{proof}

Thus, it suffices to solve the factorization problem for full and
monic linear matrices.

Let $L=A_0 + A_1x\in\Q[x]^{d\times d}$ be a full and monic linear
matrix.  Notice that $\det L\in\Q[x]$ is a univariate degree-$d$
polynomial which is not identically zero as $L$ is a full linear
matrix. Therefore, for some $i\in[d+1]$ the matrix $A_0+A_1 i$ is
invertible. Thus, replacing $x$ by $x+i$ we can assume that $A_0$ is
also invertible. We can rewrite the linear matrix as $L=(-A_0A_1^{-1}
- xI_d)(-A_1^{-1})$. Therefore,
setting $A=-A_0A_1^{-1}$, the problem is equivalent to computing the
factorization of $A-xI_d$, where $A\in\Q^{d\times d}$ is an invertible
matrix.\\

\noindent\textbf{Factorization of linear matrix $A-xI$}\\

It turns out that using standard linear algebra \cite{HK77} we can
efficiently compute a complete factorization of $A-xI_d$.

\begin{theorem}\label{lfact3}
  Given as input an invertible matrix $A\in\Q^{d\times d}$ there is an
  algorithm that computes a complete factorization of $A-xI_d$ into a
  product of linear matrix atoms in deterministic time
  $\poly(d,\bit(A))$.
\end{theorem}

\begin{proof}
  The determinant $\det (A-xI_d)$ is the characteristic polynomial
  $\chi_A(x)$ of $A$. Using the LLL algorithm we first compute
  the complete factorization of $\chi_A(x)$ over $\Q$
  \[
  \chi_A = f_1^{d_1}f_2^{d_r}\cdots f_t^{d_t},
  \]
  where each $f_i$ is a distinct irreducible factor. The algorithm works
  in two phases.\\

In this first phase, we compute the minimal polynomial
$m_A(x)\in\Q[x]$ of $A$ and also factorize it using the LLL algorithm
to get
  \[
  m_A(x) = f_1^{e_1}f_2^{e_2}\cdots f_t^{e_t}.
  \]
  By standard linear algebra \cite{HK77} each $e_i>0$.

  The algorithm computes a basis for each of the following $t$
  subspaces of $\Q^d$:
  \[
  V_i =\{v\in\Q^d\mid f_i(A)^{e_i}(v)=0\}, 1\le i\le t.
  \]
  The subspace $V_i$ consists of precisely those vectors that are
  annihilated by $f_i^{e_i}$. Since the different $f_i$ are relatively
  prime we have the following direct sum decomposition
  \[
  \Q^d = V_1 \oplus V_2\oplus \cdots \oplus V_t.
  \]
  Furthermore, since each $V_i$ is an $A$-invariant subspace, by
  choosing a basis for $\Q^d$ a union of bases for $V_1,V_2,\ldots
  V_t$, in that order, and writing the linear matrix $A-xI_d$ in that
  basis we obtain the following block diagonal form (essentially, the
  primary decomposition theorem \cite{HK77}):
\begin{equation}\label{eq-primary}
T(A-xI_d)T^{-1} =
\left(
\begin{array}{ccccc}
L_1 & 0 &0 &\ldots & 0\\
0 & L_2 &0 &\ldots & 0\\
0 & 0 &L_3 &\ldots & 0\\
& & &\ddots &\\
0 &0 &0 &\ldots & L_t\\
\end{array}
\right).
\end{equation}
The above matrix clearly factorizes as a product of $t$ linear matrices
of the form
\[
\left(
\begin{array}{ccccc}
I & 0 &0 &\ldots & 0\\
0 & I &0 &\ldots & 0\\
& \ddots&L_i &\ldots &0\\
0 & 0 & 0 &\ddots & 0\\
0 &0 &0 &\ldots & I\\
\end{array}
\right).
\]
Thus, it suffices to now consider the factorization of each linear
matrix $L_i$ which is also of the form $L_i=A_i - xI_{n_i}$, where
$n_i = d_i \cdot \deg f_i$ is the dimension of the subspace $V_i$.\\

We now describe Phase 2 of the algorithm. Notice that the
characteristic polynomial and minimal polynomial of $A_i$ are
$f_i^{d_i}$ and $f_i^{e_i}$ respectively, where $f_i$ is an
irreducible polynomial. At this point we will need some linear algebra
about the matrices whose characteristic polynomial is the power of an
irreducible polynomial.

Let $B\in\Q^{n\times n}$ be a matrix with $\chi_B=f^\ell$ and minimal
polynomial $f^e$, where $f\in\Q[x]$ is irreducible of degree $k$.
Then $\ell k =n$. We define subspaces
\[
U_j = \{u\in\Q^n \mid f^j(B) u =0\}~~ \text{ for }~~ 1\le j\le e. 
\]
By definition $U_j$ is the subspace of vectors annihilated by
$f^j(B)$. We note that
\[
U_1\subset U_2\subset \cdots U_e =\Q^n,
\]
where each $U_j$ is a proper subspace of $U_{j+1}$ for $1\le j < e$
\cite{HK77}. Furthermore, each $U_j$ is a $B$-invariant subspace
because $B\cdot g(B) = g(B) B$ for any polynomial $g$. For each $j<e$
we can alternatively describe $U_{j+1}$ as
\begin{equation}\label{eq-quotient}
U_{j+1}=\{u\in\Q^n\mid f(B) u\in U_j\}.
\end{equation}
Also,
\[
U_1 =\{u\in\Q^n\mid f(B)u=0\}.
\]
Thus $B$ restricted to $U_1$ has both minimal polynomial and
characteristic polynomial $f(x)$. Similarly, for each $j<e$ the
polynomial $f(x)$ is both the minimal and characteristic polynomial of
$B$ restricted to the quotient space $U_{j+1}/U_j, 1\le j < e$, where
the \emph{quotient vector space} $U_{j+1}/U_j$ consists of vectors of
the form $u+U_j, u\in U_{j+1}$ and $U_j$ is the zero element of the
vector space. Let $\nu_1=\dim U_1$ and $\nu_{j+1}=\dim (U_{j+1}/U_j)$.
Then $\dim U_j = \nu_1+\nu_2+\cdots + \nu_j, j\le e$.

\begin{claim}\label{u1-inv-clm}
  The $B$-invariant subspace $U_1$ is a direct sum of $\nu_1/k$ many
  $B$-invariant $k$-dimensional subspaces.
\end{claim}  

\begin{proofof}{Claim~\ref{u1-inv-clm}}
To see this claim we note that for any nonzero vector $u\in U_1$ the
so-called \emph{cyclic subspace} spanned by the \emph{cyclic basis}
$\{u,Bu,B^2u\ldots, B^{k-1}u\}$ is a $B$-invariant subspace of $U_1$
and we can repeatedly pick such subspaces until the whole of $U_1$ is
covered. Thus, $U_1$ has a \emph{good basis} which is the union of
$\nu_1/k$ many such cyclic bases, each of size $k$. With respect to
this good basis the matrix $B$ restricted to $U_1$ is block diagonal
with $\nu_1/k$ many blocks, each of size $k\times k$.
\end{proofof}

We generalize this claim to define a good basis for the quotient space
$U_{j+1}/U_j$. The matrix $B-xI$ will be easy to factorize when
expressed in terms of the basis consisting of the union of the good
bases obtained for the quotient spaces $U_{j+1}/U_j$.

\begin{claim}\label{uj-inv-clm}
  For the quotient space $U_{j}/U_{j-1}, j\ge 2$ there is a collection
  of $\nu_j/k$ pairwise disjoint sets of $k$ vectors
\[
\mathcal{B}_{ji} = \{u_{ji}, B(u_{ji}),\ldots, B^{k-1}(u_{ji})\},~~ 1\le i \le \nu_j/k
\]
such that
\begin{enumerate}
\item $\mathcal{B}_{ji}\cup U_{j-1}$ spans a subspace $U_{ji}$ of
  $U_j$ of dimension $k+\dim U_{j-1}$ for each $i$.
\item $U_{ji}\cap U_{ji'}\subseteq U_{j-1}$ for all $i\ne i'$.
\item The quotient space $U_j/U_{j-1}$ is a direct sum of the quotient
  spaces $U_{ji}/U_{j-1}$, each of which is a $k$-dimensional subspace
\item The bases $\mathcal{B_{ji}}$ can all be computed in deterministic
  polynomial time.
\end{enumerate}  
\end{claim}

\begin{proofof}{Claim~\ref{uj-inv-clm}}
  The proof is quite similar to the proof of the previous claim. For
  any vector $u\in U_j\setminus U_{j-1}$ the subset of $k$ vectors
  $\{u,Bu,\ldots,B^{k-1}u\}$ are linearly independent of $U_{j-1}$.
  Together with $U_{j-1}$ they will give a subspace of $U_j$ of
  dimension $k+\nu_{j-1}$. We can keep finding such a cyclic subsets of
  $k$ vectors as long as we have a proper subspace of $U_j$. Thus, we
  will obtain $\nu_j/k$ many such cyclic subsets $\mathcal{B}_{ji}$ as
  claimed. The construction of these bases is in deterministic
  polynomial time. As defined in the claim we have the subspaces
  $U_{ji}$ defined by these bases. Since $f$ is irreducible, any two
  distinct subspaces can intersect only in $U_{j-1}$. Thus, the
  quotient spaces $U_{ji}/U_{j-1}$ give a direct sum decomposition of
  the quotient space $U_j/U_{j-1}$. 
\end{proofof}

We define a new basis $\mathfrak{B}$ obtained by putting together the
good bases for each $U_j, 1\le j\le e$ in that order. With respect to
this basis the matrix $B-xI_\ell$ will assume the following form
\begin{equation}\label{eq-cyclic}
T_1(B-xI_\ell)T_1^{-1} =
\left(
\begin{array}{ccccc}
L'_1 & 0 &0 &\ldots & 0\\
* & L'_2 &0 &\ldots & 0\\
* & * &L'_3 &\ldots & 0\\
& & &\ddots &\\
* &* &* &\ldots & L'_e\\
\end{array}
\right)
\end{equation}
where the blocks below the diagonal blocks marked by $*$ could contain
nonzero linear forms, but the blocks above the diagonal blocks are all
zeros.  Each block $L'_j, j\ge 2$ corresponds to the quotient space
$U_j/U_{j-1}$ and, by choice of a good basis, the block $L'_j$
itself will be block diagonal with blocks of size $k$ each ($\nu_j/k$
many blocks). This yields a factorization of $B-xI_n$ as a product of
$\ell=n/k$ many linear matrix factors which are atoms by using the
following factorization repeatedly 

\begin{equation}\label{eq-factor}
\left(
\begin{array}{c|c}
A & 0 \\
\hline
D & B
\end{array}
\right) =\left(
\begin{array}{c|c}
A & 0 \\
\hline
0 & I
\end{array}
\right)\cdot
\left(
\begin{array}{c|c}
I & 0 \\
\hline
D & I
\end{array}
\right)\cdot
\left(
\begin{array}{c|c}
I & 0 \\
\hline
0 & B
\end{array}
\right).
\end{equation}
In the above equation, if the matrix on the left is a full non-unit
linear matrix then the first and third factors are full non-unit
linear matrices. The middle factor is actually a unit and can be
absorbed with either the first or the third factor.

To summarize we present below the steps of the linear matrix
factorization algorithm.

\begin{enumerate}
\item[] Input: matrix $A-xI$, where $A\in\Q^{d\times d}$ is full rank.
\item Compute the characteristic and minimal polynomials
  $\chi_A(x)=\prod_{i=1}^t f_i^{d_i}$ and $m_A(x)=\prod_{i=1}^t
  f_i^{e_i}$ of $A$ with the factorization structure as stated above.
\item[\textbf{Phase 1}] Compute bases for the subspaces
\[
V_i =\{v\in\Q^d\mid f_i^{e_i}(A)=0\}.
\]
and take their union in that order as the new basis. Then
$\Q^d=\oplus_{i=1}^t V_i$ and w.r.t.\ the new basis the matrix $A-xI$
is in block diagonal form where the $i^{th}$ block on the diagonal is
of the form $A_i - xI$ which has characteristic polynomial $f_i^{d_i}$
and minimal polynomial $f_i^{e_i}$.
\item[\textbf{Phase 2}] \textbf{for} each $1\le i\le t$ \textbf{do}
\item Call procedure GoodBasis$(A_i)$ which returns a good basis
  $\mathfrak{B}_i$ corresponding to $A_i$.
\item Find the matrix representation of $A-xI$ w.r.t.\ the basis
  $\cup_{i=1}^t \mathfrak{B}_i$.
\item Compute factorization of this matrix by repeated application
  of Equation~\ref{eq-factor}.
\item[] Procedure GoodBasis$(B)$;
  \begin{enumerate}
   \item Let $\chi_B=f^\ell$ and $m_B=f^e$, $f$ is degree $k$
     irreducible. Suppose $B$ is $n\times n$. Then $n=k\ell$.
   \item Find tower of subspaces $U_1\subset U_2\subset\cdots\subset
     U_e=\Q^n$ such that $U_j$ is annihilated by $f^j, 1\le j\le e$.
   \item Let $\mathfrak{C}_j$ be some basis for each subspace $U_j$
     computed in the above process such that $\mathfrak{C}_j\subset
     \mathfrak{C}_{j+1}$ for all $j$.
   \item $\mathfrak{C}_0=\emptyset$.  
   \item \textbf{for} each $1\le j\le e$ \textbf{do}
     \item $\mathfrak{B}_j=\mathfrak{C}_{j-1}$.
     \item \textbf{while} $\mathfrak{B}_j$ does not span $U_j$ \textbf{do}
     \item pick a vector $u\in \mathfrak{C}_j$ that is not in span of
       $\mathfrak{B}_j$ and include $\{u,Bu,\ldots,B^{k-1}u\}$ in
       $\mathfrak{B}_j$.
     \item \textbf{end-while}.
      \item \textbf{return} $\mathfrak{B}=\bigcup_{j=1}^e \mathfrak{B}_j$.
  \end{enumerate}
\end{enumerate}

\noindent\textbf{Running time analysis.}~~It is evident from the
algorithm description that the total number of field operations is
polynomially bounded in the dimension $d$ of the matrix $A$. We now
argue that the encoding sizes of the rational numbers involved in the
computation are all also polynomially bounded. The basis change matrix
$T$ used in Phase 1 (see Equation~\ref{eq-primary}) has entries of
polynomial encoding size as it is standard Gaussian elimination. In
Phase 2, the calls to Procedure GoodBasis$(A_i)$ for each $i$ are
essentially independent of each other. Thus, it suffices to analyze
one call to Procedure GoodBasis$(B)$. Again, the computation of some
basis $\mathfrak{C}_J$ for $U_j, 1\le j\le e$ is by standard Gaussian
elimination. Hence the encoding size $\bit(\mathfrak{C}_j)$ is
polynomially bounded for each $j$. The computation of the good basis
$\mathfrak{B}_j$ for the quotient space $U_j/U_{j-1}$ is done using
only $\mathfrak{C}_j$ and $\mathfrak{C}_{j-1}$. It follows that for
each $j$ we have $\bit(\mathfrak{B}_j)$ is polynomially bounded, and
hence $\bit(\mathfrak{B})$ is polynomially bounded.

\end{proof}

\subsection{The Factorization Algorithm for Matrices over $\Q[x]$}

\begin{theorem}\label{unimatfactthm}  
  Let $M\in \Q[x]^{d\times d}$ be a matrix of univariate polynomials
  over rationals where each entry of matrix $M$ is a polynomial of
  degree at most $D$. Then there is a $\poly(d,D, \bit(M))$ time
  deterministic algorithm that outputs a complete factorization of $M$
  as a product $M=M_1M_2\cdots M_r$ such that each matrix factor $M_i$
  is an atom whose entries are polynomials in $\Q[x]$ of degree at
  most $D$.
\end{theorem} 

\begin{proof}
  Given $M$ as input, we apply Higman linearization followed by the
  monicity algorithm of Lemma~\ref{monic-lem} (second part) and the linear matrix factorization algorithm of Theorem~\ref{lfact3} to
  obtain the factorization
\[
M\oplus I_s=PS'F_1F_2 \ldots F_rU'Q
\]
where each linear matrix $F_i$ is an atom, the matrix $P$ is upper
triangular with all $1$'s diagonal, the matrix $Q$ is lower triangular with all $1$'s diagonal, the matrix $S'$ is a scalar invertible matrix and $U'$ is a
unit. By absorbing $S'$ in $F_1$ and setting $U=U'Q$ we can without loss of generality assume that the factorization has the following form 
\[
M\oplus I_s=PF_1F_2 \ldots F_rU
\]

where $U$ is a unit. Moreover, the entries of $P$ and $U$ are all polynomials in
$\Q[x]$ of degree at most $D$.

We will now apply Lemma~\ref{extract} to extract the factors of $M$
(one by one from the right).

  For the first step, let $C=F_1F_2\cdots F_{r-1}$ and $D=F_rU$ in
  Lemma~\ref{extract}. The proof of Lemma~\ref{extract} yields the
  matrix $N_r=N\Pi$ such that both matrices $C''=PF_1F_2\cdots
  F_{r-1}N_r$ and $D''=N_r^{-1}F_rU$ has the first $d$ column all
  zeros except the top left $d\times d$ block of entries $c''_1$ and
  $d''_1$ which yields the nontrivial factorization $M=c''_1d''_1$,
  where $d''_1=M_r$ is an atom. Renaming $c''_1$ as $G_r$ we have from
  the structure of $C''$:
  \[
  \left( \begin{array}{cc} G_r & * \\ 0 & V_r \end{array} \right) =
  P(F_1F_2\cdots F_{r-2}) (F_{r-1}N_r).
  \]
Setting $C= F_1F_2\cdots F_{r-2}$ and $D=F_{r-1}N_r$ in
Lemma~\ref{extract} we can compute the matrix $N_{r-1}$ using which we
will obtain the next factorization $G_r=G_{r-1}M_{r-1}$, where
$M_{r-1}$ is an atom by Lemma~\ref{extract}. Note that
Lemma~\ref{extract} is applicable as all conditions are met by the
matrices in the above equation (note that the matrix $V_r$ will be a
unit).

Continuing thus, at the $i^{th}$ stage we will have
$M=G_{r-i+1}M_{r-i+1}M_{r-i+2}\cdots M_r$ after obtaining the
rightmost $i$ irreducible factors by the above process. At this stage
we will have
  \[
  \left( \begin{array}{cc} G_{r-i+1} & * \\ 0 & V_{r-i+1} \end{array}
  \right) = P(F_1F_2\cdots F_{r-i-1}) (F_{r-i}N_{r-i+1}),
  \]
  where $V_{r-i+1}$ is a unit and all other conditions are satisfied
  for application of Lemma~\ref{extract}.

  Thus, after $r$ stages we will obtain the complete factorization of the input
  matrix $M$ as
\[  
M=M_1M_2\cdots M_r,
\]
where each factor $M_i$ is an atom.\\

\subsubsection*{Running Time Analysis}

The Higman Linearization of $M$ is computed in deterministic
polynomial time. For the resulting linear matrix $L=A_0+A_1 x$, by Theorem~\ref{lfact3} its factorization as a product of
linear matrix atoms can be computed in deterministic time
$\poly(d,D,\bit(L))$.

Given factorization of $L$ into atoms, we extract atomic factorization of the input polynomial matrix $M$ as discussed above. At each stage we invoke Lemma \ref{extract} to extract an atomic factor of $M$ from right. There are total $r$ stages, $r\leq D$, where $D$ is an upper bound on the degree of polynomial entries of $M$. To see the bound $r \leq D$, note that the degree of $\det~M$ is bounded by $D$ and as for each atomic factor $M_i$ of $M$, we have degree of $\det~M_i$ at least $1$. Consequently, $r$ is upper bounded by $D$. So clearly, we can find complete factorization of $M$ into atoms in polynomially many field operations.

Now we show that the overall bit complexity of the algorithm is polynomially bounded. A crucial point to note is, the trivialization matrix $N_j$ at any stage (computed by the trivialization algorithm of Theorem \ref{triv-lem-rat}), only depends upon $C= F_1F_2\cdots F_{j-1}$, so the bit complexity of $N_j$ and $N_j^{-1}$ is polynomially bounded. Clearly, the extracted atomic factor $N_j^{-1} F_{j}N_{j+1}$ has polynomially bounded bit complexity. Moreover, the extracted factors play no role at all in the subsequent computation. This proves that the overall bit complexity of the algorithm is upper bounded by $\poly(d,D, \bit(M))$. This completes the proof of the theorem. 

\end{proof}


\bibliographystyle{alpha}
\bibliography{references}



\end{document}